\documentclass[aoas,preprint]{imsart}
\usepackage{graphicx}
\usepackage[colorlinks,citecolor=blue,urlcolor=blue]{hyperref}
\usepackage{natbib}

\arxiv{1810.05118}
\setattribute{journal}{name}{}  

\begin{document}

\begin{frontmatter}
\title{Canadian Crime Rates in the Penalty Box}
\runtitle{Canadian Crime Rates in the Penalty Box}


\author{\fnms{Simon} \snm{Demers}\corref{}\ead[label=e1]{simdem@outlook.com}\thanksref{t1}}
\thankstext{t1}{Published in {\em Journal of Community Safety and Well-Being}, {\bf 3}(3):105--109. DOI: \href{http://dx.doi.org/10.35502/jcswb.82}{10.35502/jcswb.82}. Pre-print was previously mentioned on the \emph{\href{https://www.improbable.com/2018/12/05/canadian-crime-rates-in-the-penalty-box/}{Improbable Research Blog}}.}

\runauthor{Demers}

\begin{abstract}
Over the 1962 to 2016 period, the Canadian violent crime rate has remained strongly correlated with National Hockey League (NHL) penalties. The Canadian property crime rate was similarly correlated with stolen base attempts in the Major League Baseball (MLB). Of course, correlation does not imply causation or prove association. It is simply presented here as an observation. Curious readers might be tempted to conduct additional research and ask questions in order to enhance the conversation, transition away from a state of confusion, clarify the situation, prevent false attribution, and possibly solve a problem that economists call identification.
\end{abstract}



\end{frontmatter}

\maketitle

\section{Introduction}

This short research note highlights surprisingly strong correlations between Canadian aggregate crime trends and certain professional sport trends that were observed throughout the 1962-2016 period.

Beyond any entertainment value, these empirical relationships provide new opportunities to test, validate or refute criminological theories. In the United States, for example, there appears to be a strong negative correlation between measures of consumer sentiment and rates of not only robbery and property crime \citep{RosenfeldFornango2007} but also homicide \citep{BlumsteinRosenfeld2008} and felony homicide in particular \citep{Rosenfeld2009}. This suggests that changing economic conditions can have an impact on crime. Similarly, divorce rates and crime rates also seem to move together over time \citep{Greenberg2001}, supporting the idea that major social changes and family strains can engender crime.

\section{Data}

The analysis relies on police-reported data compiled by Statistics Canada as part of its Uniform Crime Reporting (UCR) Survey. The aggregate UCR Survey data permits historical comparisons back to 1962 \citep{StatsCan2018}. Following convention, crime numbers are converted into rates per 100,000 resident population to facilitate year-over-year comparisons that take into account population growth. The focus of the analysis is on violent crime, property crime, and homicide rates.

Penalty data for the National Hockey League (NHL) was extracted directly from its official website \citep{NHLCom}. This NHL data includes the total number of penalties (Pen) and the associated number of penalty infraction minutes (PIM) for each season.

Batting data for the Major League Baseball (MLB) was downloaded from the website \cite{BaseballReferenceCom}. This MLB data includes various statistics intending to reflect offensive productivity. The analysis focuses on stolen base attempts or stolen base opportunities, which is the combination of stolen bases (SB) and "caught stealing" events (CS).


\section{Results}

\subsection{Violent crime in the NHL penalty box}

\begin{figure}[!ht]
 \caption{\label{Fig1} Canadian violent crime rate and NHL penalties, 1962-2016. Two NHL lockouts shortened the 1994-95 and 2012-13 regular seasons to 48 games instead of 82 games per team. The totals for those seasons are manually adjusted (normalized) to account for lost games and to make year-over-year comparisons possible. For its part, the 2004 NHL lockout led to a completely cancelled season. In that case, fictitious 2004 data are imputed by simply interpolating between 2003 and 2005 data points.}
 \includegraphics[width=\textwidth]{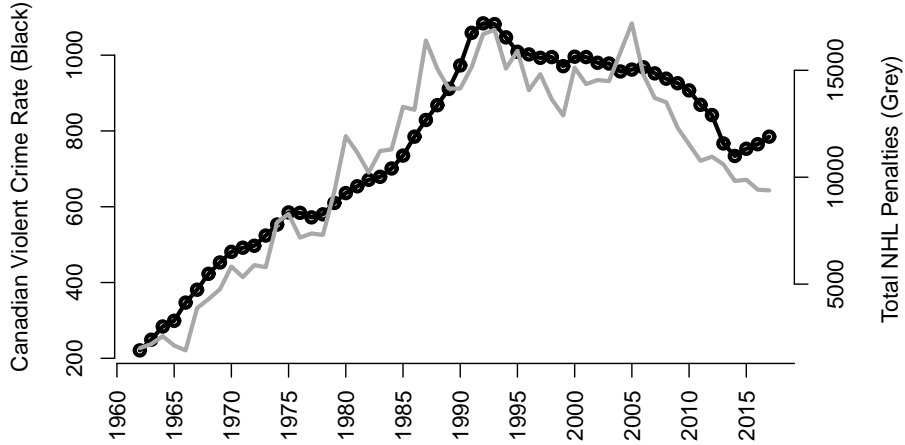}
\end{figure}

First, it turns out that the Canadian violent crime rate is positively correlated with the number of NHL penalties and penalty minutes in a surprisingly strong manner, as measured by Pearson's correlation coefficients ($r = 0.951$ and $0.925$, respectively). Fig.~\ref{Fig1} illustrates the quality of the linear fit. The ordinal association is also quite strong, as measured by Kendall's rank correlation coefficients ($\tau = 0.775$ and $0.750$, respectively).

This statistical relationship is notable because NHL penalties are typically assessed by referees for excessive toughness and physical violence and are therefore the closest thing to a violent crime in the context of professional ice hockey \citep{HeckelmanYates2003}, besides actual on-ice criminal assaults \citep{CityNews2016}.

Possible underlying factors that could drive such a strong correlation between violent crime and NHL penalties include all factors that professional hockey players would have been exposed to with essentially the same timing and same intensity as the rest of the general population. This could plausibly encompass, for example, early childhood lead exposure \citep{Nevin2000,Nevin2007,Wolf2014} and a social civilizing or pacifying process \citep{Spierenburg2001,Stille2003,Restrepo2015}.

The latter argument is a sociological one. It is based on the idea that society in general may have become more enlightened over the last few decades. Better self-control and social pressures would therefore explain parallel reductions in both crime and on-ice penalties. The former argument is built on the idea that hockey players who played in the NHL throughout the late 1980s and early 1990s may have been exposed to harmful amounts of lead when they were growing up in the 1960s and 1970s, \emph{before} leaded gasoline and lead-based paint became regulated \citep{Wolf2014}. In contrast, hockey players born after the mid-1970s were more likely to grow up with healthier brains and a lower propensity for violence. Canadians and Canadian youths in particular would have been similarly affected.

It is not possible at this stage to completely eliminate the possibility that NHL penalties might actually stimulate violent crime at the national level, through some kind of copycat or emulation effect for instance \citep{Moser2004,Adubato2011,CardDahl2011,Kirbyetal2014}. Although this may sound far-fetched at first glance, earlier studies did reveal some correlation between the success of sports teams and reductions in suicide \citep{Joineretal2006} and even homicide rates \citep{Fernquist2000}. In fact, some studies examined the relationship between homicide rates and the introduction of television in certain countries and found a consistently positive correlation between the two \citep{Centerwall1989}, revealing that not only televised sport -- but television more generally -- has the potential to engender violence in society \citep{Centerwall1992}.

We note in passing that the correlation between violent crime and NHL penalties is unlikely to be explained away by annual weather or temperature variations \citep{Burkeetal2015} because NHL games are generally played in a climate-controlled environment, unlike baseball \citep{Reifmanetal1991,Larricketal2011} or American football games \citep{Craigetal2016} for instance.

\subsection{Property crime around the baseball diamond}

\begin{figure}[!ht]
 \caption{\label{Fig2} Canadian property crime rate and stolen base attempts in the MLB, 1962-2016. Three significant MLB strikes resulted in lost games during the 1972, 1981, 1994, and 1995 seasons. However, the MLB data are averaged on a per-game basis, which removes the need to normalize the data or simulate full seasons in order to account for lost games.}
 \includegraphics[width=\textwidth]{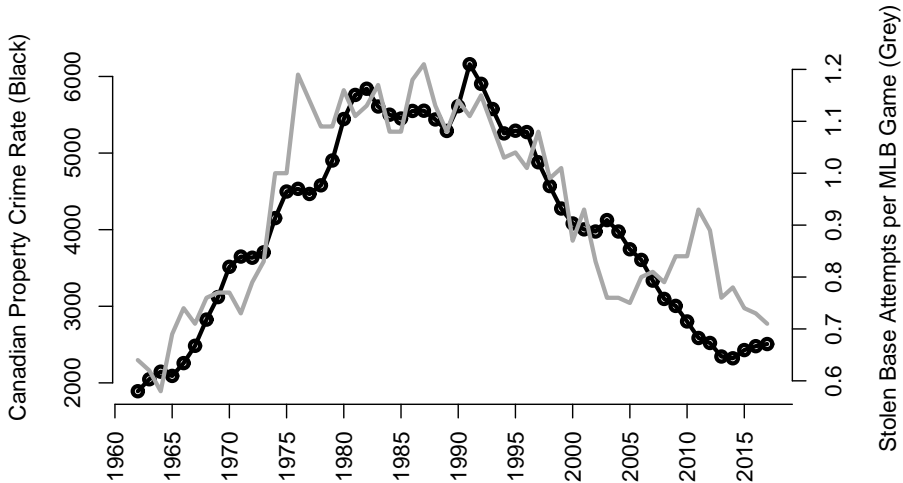}
\end{figure}

Second, it turns out that the Canadian property crime rate is positively correlated with the rate of stolen base attempts in the MLB ($r = 0.894$ and $\tau = 0.696$). Fig.~\ref{Fig2} illustrates the quality of the linear fit.

Stolen base attempts reflect the propensity of baseball players to opportunistically try to move forward around the bases at the expense of the defensive team, in order to possibly score more runs as a result. Stolen base opportunities can therefore properly be seen as the professional baseball equivalent of a \emph{lawful} property crime.

Plenty of explanations have been professed to explain the relative rise of base stealing in the mid-1970s \citep{McMurray2015}, its peak throughout the 1980s \citep{James1982}, and its decline in the mid-1990s \citep{Kurkjian1994}. However, identifying the underlying factors that could explain why stolen bases in the MLB may be related to property crime in Canada remains an open problem.

\begin{figure}[!ht]
 \caption{\label{Fig3} Canadian homicide rate and stolen base attempts in the MLB, 1962-2016. The Canadian homicide rate is actually correlated most strongly with \emph{failed} stolen base attempts where the runner was caught stealing ($r = 0.822$), while the rate of attempted murders is actually correlated most strongly with \emph{successful} stolen bases ($r = 0.906$). This may look incongruent at first glance, but this is only because stolen bases are defined from the point of view of the runner. From the point of view of the defensive team, a runner who is caught stealing represents a \emph{successful} out.}
 \includegraphics[width=\textwidth]{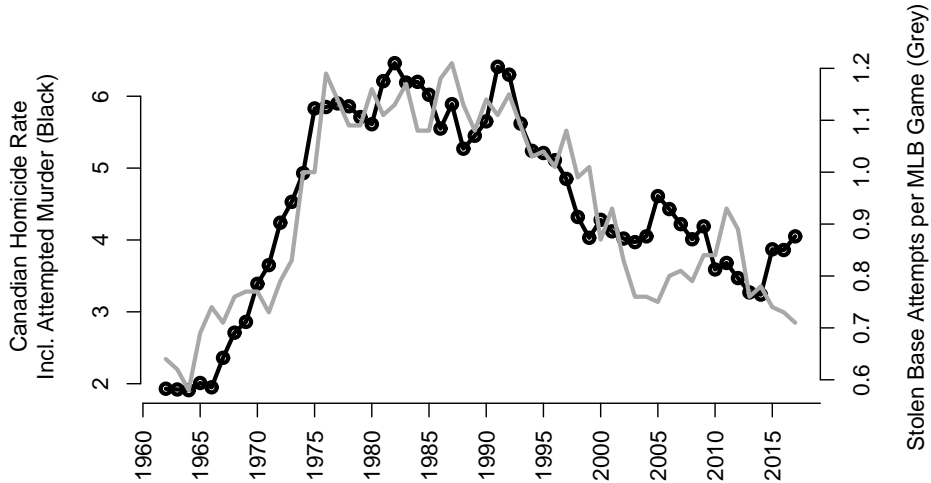}
\end{figure}

As a side note, \cite{Farrelletal2018} already highlighted that there was a remarkable similarity between historical property crime and homicide rates in Canada, once attempted murders and actual homicides are aggregated into a single measure of total homicides ($r = 0.91$). In that context and in light of the previously discussed results, it should not come as a surprise that Canadian homicides and attempted murders, like property crimes, are also correlated with stolen base attempts ($r = 0.894$ and $\tau = 0.695$). This is shown by Fig.~\ref{Fig3}.

\cite{Farrelletal2018} speculated that similarities between historical homicide and property crime rates may have been driven by the progressive implementation of security improvements, especially in the second half of the 1990s. While it is not immediately obvious whether (or how) this security hypothesis may also apply to stolen base opportunities in the MLB, more research into this area could be enlightening (and probably entertaining).

\section{Conclusion}

Correlation does not imply causation and it is entirely possible that the statistical relationships presented in this short research note may be coincidental or spurious, in the vein of \cite{Andersen2012} and \cite{Vigen2015} for example. However, they also very well could be examples of life imitating sports \citep{Steen2014} or vice versa \citep{Goldstein1989}.


More research would obviously be required to assess whether these statistical relationships are purely coincidental, to determine to what extent both trends might be influenced by common underlying factors, and to investigate the possibility that they might be fueled endogenously through a copycat effect, among many other possible pathways.

\end{document}